\documentclass[12pt]{article}
\usepackage[dvips]{graphicx}
\usepackage{epsfig}
\usepackage{amsmath,amssymb,amsbsy}
\usepackage{amsfonts}
\usepackage{bm}
\newcommand{\be}{\begin{equation}}
\newcommand{\ee}{\end{equation}}
\newcommand{\beq}{\begin{eqnarray}}
\newcommand{\eeq}{\end{eqnarray}}
\newcommand{\no}{\nonumber}
\begin{document}
\thispagestyle{empty}
\begin{center}

{\Large\bf{The transverse momentum dependent

\vskip 0.4cm
 statistical parton distributions revisited}}

\vskip1.4cm
{\bf Claude Bourrely}
\vskip 0.3cm
 Aix-Marseille Universit\'e,
D\'epartement de Physique,\\ Facult\'e des Sciences de Luminy,
13288 Marseille, Cedex 09, France\\
\vskip 0.5cm
{\bf Franco Buccella}
\vskip 0.3cm
INFN, Sezione di Napoli, Via Cintia, Napoli, I-80126, Italy
\vskip 0.5cm
{\bf Jacques Soffer}
\vskip 0.3cm
Physics Department, Temple University,\\
Barton Hall, 1900 N, 13th Street\\
Philadelphia, PA 19122-6082, USA
\vskip 0.5cm
{\bf Abstract}\end{center}

The extension of the statistical parton distributions to include
their transverse momentum dependence (TMD) is revisited by considering that the proton target has a finite longitudinal momentum. The TMD will be generated by means of a transverse energy sum rule. The new results are mainly relevant for electron-proton inelastic collisions in the low $Q^2$ region. 
We take into account the effects of the Melosh-Wigner rotation for the helicity distributions.

\vskip 0.5cm

\noindent {\it Key words}: Parton distributions; Statistical approach; Helicity; Melosh-Wigner rotation\\
\noindent PACS numbers: 12.40.Ee, 13.60.Hb, 13.88.+e, 14.65.Bt
\vskip 0.5cm

\vskip 0.5cm
\section{Introduction}

In 2002 \cite{BBS1} we have proposed a description of the parton distributions
inspired by quantum statistics with some robust phenomenological motivations.

1) The defect in the Gottfried sum rule \cite{Gott,NMC}, implies $\bar d(x) > \bar u(x)$,
 advocated many years ago as a consequence of the Pauli principle \cite{Pauli}. This
inequality has been confirmed by Drell-Yan production of muon pairs \cite{DY}, up to some moderate
$x$ values and remains to be verified for $x>0.2$ or so.

2) The dramatic decrease at high $x$ of the ratio of the structure functions $F^n_2(x)/F^p_2(x)$ \cite{SVS}, strongly connected to
the behavior of $d(x)/u(x)$, which is predicted to flatten out for $x>0.6$ \cite{BBS3}.

3) The fast increase at high $x$ of the double longitudinal-spin asymmetry
 $A_1^p(x)$, which implies the dominance, in that region, of the $u$-quark
with the helicity along the polarization of the proton target.\\
The correlation between the shapes and the first moments dictated by the Pauli 
principle, allowed us to describe with a small number of parameters both 
the unpolarized and polarized distributions with the important consequence 
to predict a positive value for $\Delta \bar{u}(x)$ and a negative one for 
$\Delta \bar{d}(x)$, giving a positive contribution to the Bjorken sum rule \cite{BJ}.\\
Typical predictions of our approach, as the monotonic increase of the positive ratio 
$\Delta u(x)/u(x)$ and decrease of the negative ratio $ \Delta d(x)/d(x)$, are in a good agreement with
experiment \cite{hermes,JLab}. It should be stressed that our prediction on $ \Delta 
d(x)/d(x)$ does not correspond to the belief that this quantity, according 
to the counting rules, should change sign to reach +1 for $x = 1$.

In earlier works \cite{BBS4,BBS5}, we succeeded to explain the arbitrary factors, which were 
necessary to agree with data by the extension to the transverse degrees 
of freedom and we constructed a set of transverse momentum dependent (TMD) statistical parton distributions. 
Here we want to improve our approach by considering the longitudinal momentum of the proton target $P_z$ finite and not $\infty$, as it is only
in the limit $Q^2 \rightarrow \infty$.\\
The paper is organized as follows. In the next section we review the construction of the statistical distributions and then we derive the exact transverse energy sum rule, the TMD statistical distributions and the expression of $P_z$ as a function of $Q^2$ and $x$. In section 3 we recall the expression of the Melosh-Wigner transformation which has an important role for the TMD helicity distributions, as will be shown in section 4. We give our concluding remarks in section 5. A useful transformation to simplify the calculation of integrals over the transverse momentum of the partons is
given in the Appendix.

\newpage
\section{The statistical approach}
In our previous papers \cite{BBS1,BBS3} we took the following expression for
 the {\it nondiffractive} contribution
of the quark distributions $q^{h}(x)$ of flavor $q$ and helicity $h$, at an
input energy scale $Q_0^2$
\be
x q^{h}(x) = \frac{A X_{0q}^{h} x^b}
{\mbox{exp}[(x - X_{0q}^{h})/{\bar x}] +1} \,,
\label{eq1}
\ee
where $X_{0q}^{h}$ plays the role of a {\it thermodynamical potential} and
$\bar x$ of the {\it universal temperature}.
Correspondingly for antiquarks we took the expression 
\be
x \bar q^{h}(x) = \frac{\bar A  x^{\bar b}}
{X_{0q}^{-h}\left[{\mbox{exp}[(x + X_{0q}^{-h})/{\bar x}] +1}\right]} \,,
\label{eq2}
\ee
which shows a strong connection between quarks and antiquarks of opposite
helicity. In order to take into account the uniform rapid rise of all distributions in the very low $x$ region, we had to add to the above nondiffractive contributions, for quarks and antiquarks, the following {\it diffractive} contribution
\be
x q^{D}(x) = \frac{\tilde A  x^{\tilde b}}
{\mbox{exp}(x/\bar x) +1} \,,
\label{eq2d}
\ee
which is flavor and helicity independent.\\
 Finally the expression for the gluon 
distribution was
\be
xG(x) = \frac{A_G x^{b_G}}{\mbox{exp}(x/\bar x) -1}\,,
\label{eq3}
\ee
since the potential of the gluon $X_{0G}$ must be zero.\\
{}From the well-established features of the $u$ and $d$ quark distributions,
extracted
from Deep Inelastic Scattering (DIS) data, we could anticipate some simple relations
between
the corresponding potentials, namely
\beq
X_{0u}^{+}& >& X_{0u}^{-} \label{eq4} \\
X_{0d}^{-}& >& X_{0d}^{+} \label{eq5}\\
X_{0u}^{+} + X_{0u}^{-}& >& X_{0d}^{+} + X_{0d}^{-} ~.\label{eq6}
\eeq
Clearly these partons distributions $p_i= q, \bar {q}, G$ must obey the
momentum sum rule, which reads
\beq
\sum_i \int_0^1 x p_i(x) dx &=& 1~. \label{eq7}
\eeq
To  account for the factors $ X_{0q}^{h}$ in  the numerator of Eq.(\ref{eq1})
and $ X_{0q}^{-h}$
in the denominator of Eq.(\ref{eq2}), we extended our analysis to the
transverse
degrees of freedom \cite{BBS4}. If $p_T$ stands for the parton transverse
momemtum, the corresponding TMD statistical distribution
$p_i(x,p_{T}^2)$ is normalized such that $\int p_i(x,p_{T}^2) dp_{T}^2 =
p_i(x)$. Let us denote by $E$ the energy of the proton target of mass $M$ and
longitudinal momentum $P_z$. By definition the $z$ axis is along the direction of the proton momentum and this implies its transverse momentum
$P_T$ to be zero. The deep inelastic regime, where parton model may be applied, is at high $Q^2$ and in that limit, one may neglect $M$ with respect to $P_z$. However at very small $x$, one may not neglect $p_T$ with respect to $xP_z$. The proton
energy is taken to be $E=P_z + M^2/2P_z$. Similarly the energy of a massless parton, such
that $xP_z >> p_T$, is $xP_z + p^2_T/2xP_z$. 
So by writing a sum rule for the energy of the partons analogous
to (\ref{eq7}), we see that it does not depend on $P_z$ and one gets now
\begin{equation}
M^2 = \sum_i \int_0^1dx \int_0^\infty p_i(x,p^2_T)\frac{p^2_T}{x}dp^2_T~.
\label{eq8}
\end{equation}
As explained first in Ref.\cite{BBS4} and more recently in Ref.\cite{BBS5}, an
improved version, this energy sum rule implies the $p_{T}^2$ dependence of the nondiffractive contribution of the quark distributions to be
\be
\frac{1}{\mbox{exp}(p^2_{T}/x \mu^2 - Y_{0q}^h) +1} \,,
\label{eq9}
\ee
where $Y_{0q}^h$ is the thermodynamical potential associated to the parton
transverse momentum $p_T$ and
$1/\mu^2$ is a Lagrange multiplier, whose value is determined by the transverse
energy sum rule.\\
Before moving on, we would like to make an important remark on the implication of the above transverse energy sum rule.
If one makes the commonly used assumption that $p_i(x,p_T^2)$ obeys a factorization property, such as, $p_i(x,p_T^2) = p_i(x)\cdot f_i(p_T^2)$, then Eq.(\ref{eq8}) implies that $\int_0^1dx~p_i(x)/x$ must converge. 
Consequently for all partons, the small $x$ behavior is strongly constrained, since it must vanish like $x^{\alpha_i}$, with $\alpha_i >0$. This is in obvious disagreement with the observed rapid rise of sea quarks and gluon in the very small $x$ region. We conclude that our transverse energy sum rule does not allow the simplifying factorization assumption for the TMD distributions.

\subsection{Transverse energy sum rule and $p_T$ dependence}

The approximations $P_z >> M$ and $xP_z >> p_T$, valid in the limit $P_z
=\infty$, bring in an unphysical singularity for $x \rightarrow 0$, where 
evidently $xP_z$ is not larger than $p_T$.
We have the relations
\beq
E &=& \sum_i E_i \label{eq8a} \\
E -P_z &=& \sum_i(E_i -p_{zi})   \label{eq9a} \,,
\eeq
where $p_{zi} = x_{i}P_z$,
$E_i = \sqrt{x_{i}^2P_z^2 +p_{Ti}^2 +m_i^2}$ is the energy of quark $i$, 
and $E = \sqrt{P_z^2 + M^2}$. From now on we shall neglect the quark mass $m_i$. By multiplying both sides of Eq.(\ref{eq9a}) by 
$E +P_z$ and inside the summation, both numerator and denominator, by $E_i +p_{zi}$, we get
\beq 
M^2 &=& (E +P_z) \sum_i \frac{(E_i -P_i)(E_i +p_{zi})}{(E_i +p_{zi})} 
p_i(x_i,p^2_{Ti})\label{eq10a}\\
&=& (E +P_z)\sum_i \frac{p_{Ti}^2}{\sqrt{x_{i}^2P_z^2 +p_{Ti}^2} 
+x_iP_z}p_i(x_i,p^2_{Ti})~.
\label{eq10b}
\eeq
Therefore instead of Eq.(\ref{eq8}) if we consider Eq.(\ref{eq10b}) in the continuum 
limit, the exact transverse energy sum rule reads
\beq
M^2\!\!\!\ &=&\!\!\!\!(\sqrt{P_z^2 + M^2} + P_z) \sum_i \int_0^1dx \int_0^\infty
p_i(x,p^2_T)\frac{p^2_T dp^2_T }{\sqrt{x^2P_z^2 + p_T^2} + xP_z}
\label{13a}\\
&\simeq& \sum_i \int_0^1dx \int_0^\infty
p_i(x,p^2_T)\frac{2p^2_T dp^2_T}{\sqrt{x^2 + p_T^2/P_z^2} + x} \,,
\label{eq13}
\eeq
since $P_z >> M$.\\

By comparing the integrants of Eq.(\ref{eq8}) and of Eq.(\ref{eq13}), we
see that the $p_T$ dependence of the parton distribution, 
which was driven in Eq.(\ref{eq9}) by $p_T^2/x\mu^2$, 
is now replaced by the following expression involving $P_z$
\be
\frac{1}{\mbox{exp}[2p^2_{T}/[\mu^2(x + \sqrt{x^2 + p_T^2/P_z^2})] - Y_{0q}^h]
+1} \,.
\label{eq15}
\ee
As a result the TMD of the nondiffractive contribution of the quark distributions are now given by the following expression
\beq
xq^h(x,p_T^2,P_z) &=& \frac{1}
{\mbox{exp}[(x - X_{0q}^{h})/{\bar x}] +1}\times\no \\
&&\frac{F_q^h(x,P_z,\mu^2)}
{\mbox{exp}[2p^2_{T}/
[\mu^2(x + \sqrt{x^2 + p_T^2/P_z^2} )] - Y_{0q}^h]
+1}\,,\label{eqF}
\eeq
where $F_q^h$ is a normalization function , which has the dimension of the inverse of
an energy square, and which will be explained in more details below (see Section 4).

\subsection{Expression for $P_z$ finite}
The value of $P_z$ for finite $Q^2$, which seems more natural to us, is 
its value in the reference frame where the final hadrons are at rest. In this 
reference frame the proton target and the virtual photon have opposite
momentum. To go to this frame from the lab frame where the proton is at rest, we have therefore to consider the 
Lorentz transformation with velocity
\be
v = \frac{|\overrightarrow{q}|}{q_0 +M}\,,
\label{eq16}
\ee
where $(q_0, \overrightarrow{q})$ is the virtual photon momentum.
In that reference frame
\beq
|P_z| = \frac{M \frac{|\overrightarrow{q}|}{q_0 +M}}
{\sqrt{1 - \frac{|\overrightarrow{q}|^2}{(q_0+M)^2}}} 
 = \frac{M |\overrightarrow{q}|}
{\sqrt{q_0^2 +2 Mq_0 + M^2 -|\overrightarrow{q}|^2}} \,,
\label{eq17}
\eeq
which implies 
\be 
P_z^2 = \frac{M^2 (Q^2 +q_0^2)}{2 M q_0 +q_0^2 +M^2 -|\overrightarrow{q}|^2}
\,.
\label{eq18}
\ee
Now from the definition of the scaling variable $x = Q^2/2P\cdot q$, we get
\be
Q^2=2x M q_0~,
\label{eq19}
\ee
so we can write
\beq
P_z^2 = \frac{M^2 (Q^2 +\frac{Q^4}{4x^2 M^2})}
        {\frac{Q^2}{x} + M^2 -Q^2} 
 = \frac{x M^2 (\frac{Q^2}{4x^2 M^2} +1)}
{1 -x +\frac{xM^2}{Q^2}} \,.
\label{eq20}
\eeq
Finally we get the following simple expression of $P_z$ in terms of  $x$
and $Q^2$
\beq
P_z = \sqrt{\frac{\frac{Q^2}{4x} + xM^2}{1 -x +\frac{xM^2}{Q^2}}} \,.
\label{eq21}
\eeq
\begin{figure}[ht]
\begin{center}
  \epsfig{figure=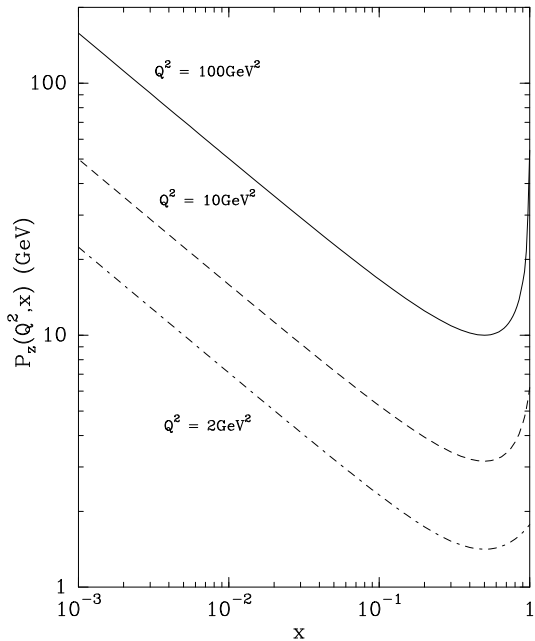,width=7.0cm}
\end{center}
 \vspace*{-5mm}
\caption{
$P_z$ versus $x$ for different $Q^2$ values.}
\label{pz}
\vspace*{-1.5ex}
\end{figure}
For illustration we diplay in Fig. 1, $P_z$ versus $x$ for different $Q^2$ values.
We notice that when $Q^2 =\infty$, then $P_z =\infty$ if $x \neq 0$. However for
$Q^2$ finite, when $x \rightarrow 0$, we also have $P_z \rightarrow \infty$.\\
 From Eq.(\ref{eq21}) one realizes that the effect of the finite value for $P_z$
 in Eq.(\ref{eqF}) is relevant at high values of $p^2_T$, for $x$ in the neighborough of 1 and for small values of
 $Q^2$.
\section{The Melosh-Wigner transformation}
So far in all our quark or antiquark TMD distributions, the label "`$h$"' stands for the
helicity along the
longitudinal momentum and not along the direction of the momentum, as normally
defined for a genuine helicity. The basic effect of a transverse momentum $p_T
\neq 0$ is the Melosh-Wigner rotation \cite{mel-wig,bucc}, which mixes the
components $q^{\pm}$ in the following way
\begin{equation}
q^{+MW}= \cos^2\theta ~q^+ + \sin^2\theta ~q^- ~~~~\mbox{and}~~~q^{-MW}=
\cos^2\theta ~q^- + \sin^2\theta ~q^+,
\label{mel1}
\end{equation}
where, for massless partons,
\be
\theta = \arctan{(\frac{p_T}{p_0 +p_z})}~,
\label{me1}
\ee
with $p_0 = \sqrt{p_T^2 +p_z^2}$.\\

Consequently $q = q^+ + q^-$ remains unchanged since $q^{MW}=q$, whereas we have
\begin{equation}
\Delta q^{MW}= (\mbox{cos}^2\theta - \mbox{sin}^2\theta) \Delta q~.
\label{mel2}
\end{equation}
The angle of the Melosh-Wigner transformation Eq.(\ref{me1}), originally derived in Ref. \cite{mel-wig}, has the correct property \footnote{The angle we used in Ref. \cite{BBS5}, which was phenomenological, didn't have this property.} to vanish
when either $p_T = 0$ or $P_z \rightarrow \infty$.

From simple calculations \footnote{ This result has been also used in several papers on the proton spin puzzle \cite{MZ,MSS}.} we get
\beq
&&\mbox{cos}^2\theta - \mbox{sin}^2\theta
 = \frac{(p_0 +p_z)^2 -p_T^2}
{(p_0 +p_z)^2 +p_T^2} = \frac{p_z}{p_0}
= \frac{x P_z}{\sqrt{x^2 P_z^2 +p_T^2}} \no \\
&& =\frac{1}{\sqrt{1 +\frac{p_T^2}{x^2 P_z^2}}} =
\frac{1}{\sqrt{1 +\frac{\xi \mu^2}{x P_z^2}}} =
\frac{1}{\frac{2\xi}{\eta} -1} \,,\label{eq239}
\eeq
where we have used the variables $\xi$ and $\eta$ introduced in the Appendix and Eq.(\ref{eq28}).

In the last term of Eq.(\ref{eq239}) by using the expression of $\xi$ given in Eq.(\ref{xidef}), we
 obtain $ 1 /(1+ \frac{\mu^2 \eta}{2 x P_z^2})$ and 
this term cancels $1+ \frac{\mu^2 \eta}{2 x P_z^2}$ in
Eq.(\ref{eq222}).\\

The consequences of this cancellation will be discussed in the next section.

\section{Some explicit expressions for the TMD distributions}
One of the goals of this section is to show how to use the normalization
function $F_q^h$ introduced in Eq.(\ref{eqF}) and then how to evaluate it.

Following Eq.(\ref{eq22}) and the results of the Appendix, the nondiffractive contribution of the quark distributions $q^h(x)$ read
\be
\int_0^\infty xq^{h}(x,p_T^2,P_z)dp_T^2 =
\frac{F_q^h(x,P_z,\mu^2)}
{\mbox{exp}[(x - X_{0q}^{h})/\bar x] +1}\mu^2 x\ln{(1+e^{Y_{0q}^{h}})}[1 + R_q^h]
\label{eq77}\,.
\ee
For illustration we display in Fig. 2, $R_u^+$, using the parameters $\mu^2=0.198\mbox{GeV}^2$ and $Y_{0u}^+ =1.122$, determined in Ref. \cite{BBS5}. The other correction factors $R_u^-$, $R_d^+$, $R_d^-$ are very similar, so it is clear that
the effect of these correction factors is limited to the small $Q^2$ region.
\begin{figure}[htb]
\begin{center}
  \epsfig{figure=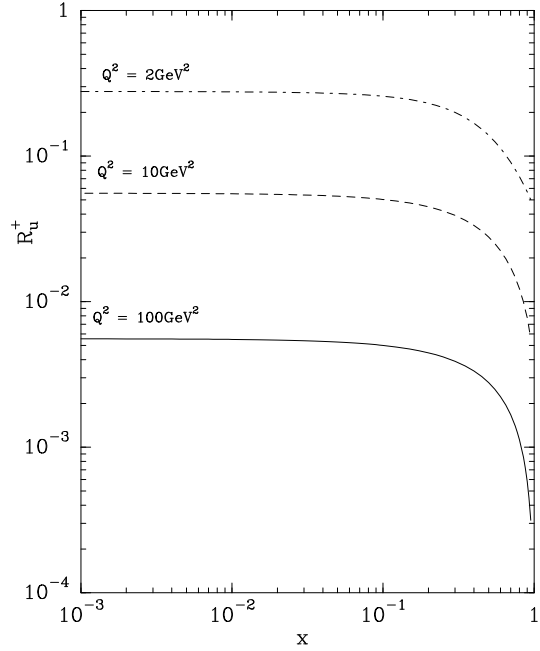,width=7cm}
\end{center}
  \vspace*{-5mm}
\caption{
$R_u^+$ versus $x$ for different $Q^2$ values.}
\label{R}
\end{figure}
By using Eq.(\ref{eq1}) one gets the following condition for $F_q^h$
\be
F_q^h(x,P_z,\mu^2) = \frac{A X_{0q}^{h} x^b }
{\mu^2 x\ln{(1+e^{Y_{0q}^{h}})}[1 + R_q^h]}
\label{eq110}\,.
\ee
The parameters $A,~b,~X_{0q}^h$
are known, $P_z$ is a function of $Q^2,~x$, the only parameters to be fixed
are $\mu^2$ and $Y_{0q}^h$. It is worth noting that since $q^h(x)$ should not depend
on $P_z$, $F_q^h$ is such that it will absorb the $P_z$ dependence occuring only in the
correction factor $R_q^h$.\\
The unpolarized TMD distributions are defined as
\beq
&&xq(x,p_T^2,P_z) = \no\\
&&\!\!\frac{1}
{\mbox{exp}[(x - X_{0q}^{+})/{\bar x}] +1}\
\frac{F_q^+(x,P_z,\mu^2)}
{\mbox{exp}[2p^2_{T}/
[\mu^2(x + \sqrt{x^2 + p_T^2/P_z^2} )]\! -\! Y_{0q}^+]
\!+\!1}\no\\
&&\!\!+\frac{1}
{\mbox{exp}[(x - X_{0q}^{-})/{\bar x}] +1}
\frac{F_q^-(x,P_z,\mu^2)}
{\mbox{exp}[2p^2_{T}/
[\mu^2(x + \sqrt{x^2 + p_T^2/P_z^2} )]\! -\! Y_{0q}^-]
\!+\!1} \label{eq111a}\,.
\eeq
For illustration we display in Fig. 3, the TMD of the nondiffractive contributions of the quark distributions $xu(x,p_T^2,Q^2)$ and $xd(x,p_T^2,Q^2)$,
using the above normalization condition Eq.(\ref{eq110}) and the parameters of Ref.\cite{BBS5}. It is clear that all these $p_T$ distributions are close to a Gaussian behavior, but with a $x$-dependent width. As expected, the fall off in $p_T$ is faster for smaller $x$ values and we see that the effect of the correction factor $R_q^h$, which decreases the distributions, is more important near $p_T = 0$.

\begin{figure}[ht]
\begin{center}
  \begin{minipage}{6.5cm}
  \epsfig{figure=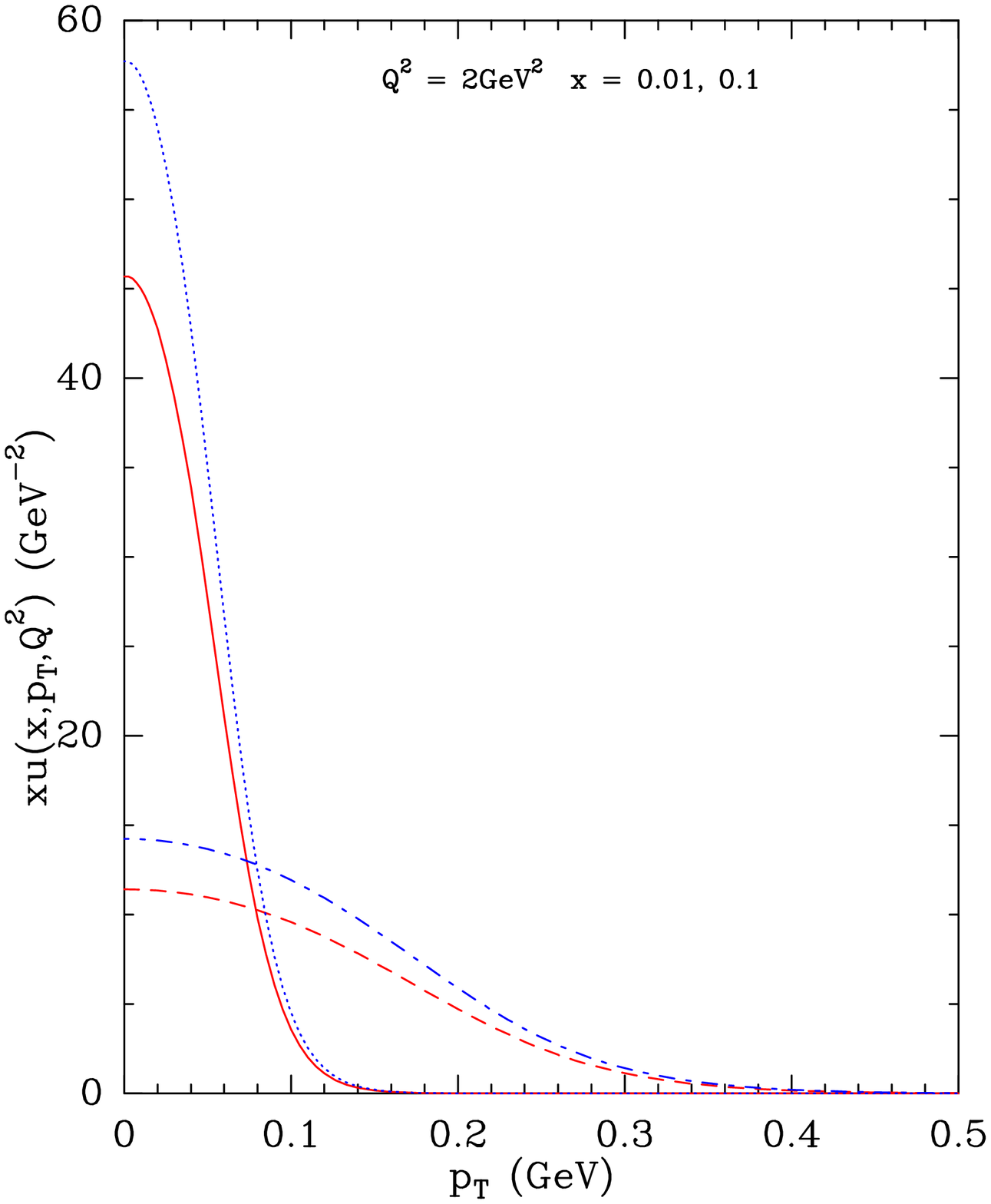,width=6.5cm}
  \end{minipage}
    \begin{minipage}{6.5cm}
  \epsfig{figure=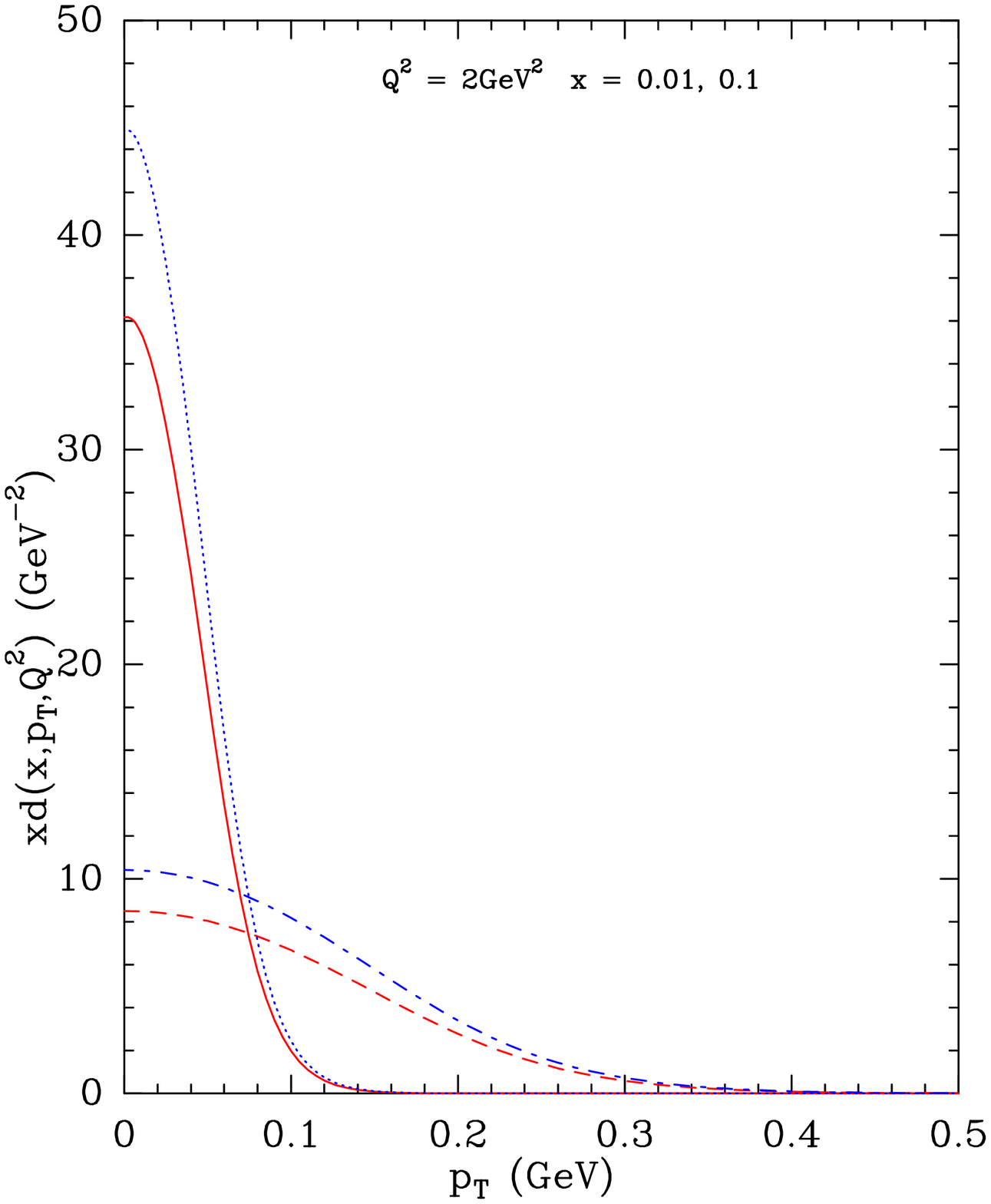,width=6.5cm}
    \end{minipage}
\end{center}
  \vspace*{-5mm}
\caption{The TMD nondiffractive contribution of the quark distributions versus $p_T$, at $Q^2=2\mbox{GeV}^2$, for two $x$ values: solid and dotted lines $x=0.01$, dashed and dot-dashed lines $x=0.1$. (\it{upper curves $R_q^h = 0$, lower curves $R_q^h \neq 0$})}
\label{dsigcross}
\vspace*{-1.5ex}
\end{figure}
Let us now turn to the diffractive contribution Eq.(\ref{eq2d}). Since $\tilde b <0$, one cannot introduce the $p_T$ dependence
similarly to the nondiffractive contributions, because it generates a singular behavior in the energy sum rule, when $x \to 0$. In order to avoid this difficulty, as in Ref.\cite{BBS5}, we modify our prescription by taking
\beq
x q^D(x,p_T^2,P_z) &=& \frac{1}
{\mbox{exp}[x/{\bar x}] +1}\times \no \\
&&\frac{ F^D(x,P_z,\mu^2)}
{\mbox{exp}[2p^2_{T}/
[\mu^2 x(x + \sqrt{x^2 + p_T^2/P_z^2} )]]
+1}\,,\label{eq3diff}
\eeq
whose $p_T$ fall off is stronger, because $\mu^2$ is now replaced by $x\mu^2$.
In order to make a link with Eq.(\ref{eq2d}), we have to compute the integral
\be
 x q^D(x) =\int_0^\infty x q^D(x,p_T^2,P_z)dp_T^2\,.
\label{eq4diff}
\ee
By using Eq.(\ref{eq112}) with the substitutions $Y^h_{0q}=0$ and $\mu^2 \to x\mu^2$, we easily deduce from Eq.(\ref{eq4diff})
\be
F^D(x,P_z,\mu^2) = \frac{\tilde A  x^{\tilde b-2} }
{\mu^2 [\ln2 + \frac{\pi^2 \mu^2}{24P_z^2} ]}
\label{eq10diff}\,,
\ee
since $\mbox{Li}_2(-1) = -\pi^2/12$. This result is similar to what was found in Ref.\cite{BBS5}, because the correction factor is very small.

Concerning the TMD gluon distribution, it reads similarly
\beq
xG(x,p_T^2,P_z) &=& \frac{1}
{\mbox{exp}[x/{\bar x}] -1}\no \\
&&\frac{F^G(x,P_z,\mu^2)}
{\mbox{exp}[2p^2_{T}/
[\mu^2 x(x + \sqrt{x^2 + p_T^2/P_z^2} )] + Y_G]
-1}\,.\label{eq3g}
\eeq
Here we must introduce a negative potential $-Y_G$ to avoid a singularity when
$Y_G \rightarrow 0$.
In order to make a link with Eq.(\ref{eq3}) we have to compute the integral
\be
 xG(x) =\int_0^\infty xG(x,p_T^2,P_z)dp_T^2\,,
\label{eq4g}
\ee
which reduces to
\beq
\int_0^\infty \frac{dp^2_{T}}
{\mbox{exp}[2p^2_{T}/
[\mu^2 x(x + \sqrt{x^2 + p_T^2/P_z^2} )] + Y_G]-1}&=& \no \\ 
\int_0^{\infty} \frac{\mu^2 x (1 +\frac{\mu^2 \eta}{2 x P_z^2})   d\eta}
{\mbox{exp}[ \eta +Y_G] -1}\,. \label{eq5g}
\eeq
and by using Eq.(\ref{eq4g}) one obtains
\beq
&&F^G(x,P_z,\mu^2) = \label{eq10g} \\
&&\frac{A_G x^{b_{G}-2} }
{\mu^2 \left[-\ln{(1-e^{-Y_G})}\!
+\!\frac{\mu^2}{2P_z^2} [\frac{1}{6}\pi^2  
-\frac{1}{2}Y_G^2  +Y_G\ln{(e^{Y_G}-1)}+\mbox{Li}_2(1-e^{Y_G})]\right]}
\no\,.
\eeq
This result is similar to what was found in Ref.\cite{BBS5}, because the correction factor is very small since we take $Y_G =10^{-6}$.\\
Before we move on, we would like to compare our results with the relativistic covariant approach \cite{ESTZ}, where they introduce the
variable $x + p_T^2/xM^2$ combining the $x$ and $p_T$ dependences. We show in Fig. 4 the result of our calculations which can be compared
with the results of Ref.\cite{ESTZ} displayed in their Fig. 1. The two results are compatible with a broader shape for increasing $x$, but one notices that the $p_T$ fall off is less rapid in our case.

\begin{figure}[ht]
\begin{center}
  \begin{minipage}{6.5cm}
  \epsfig{figure=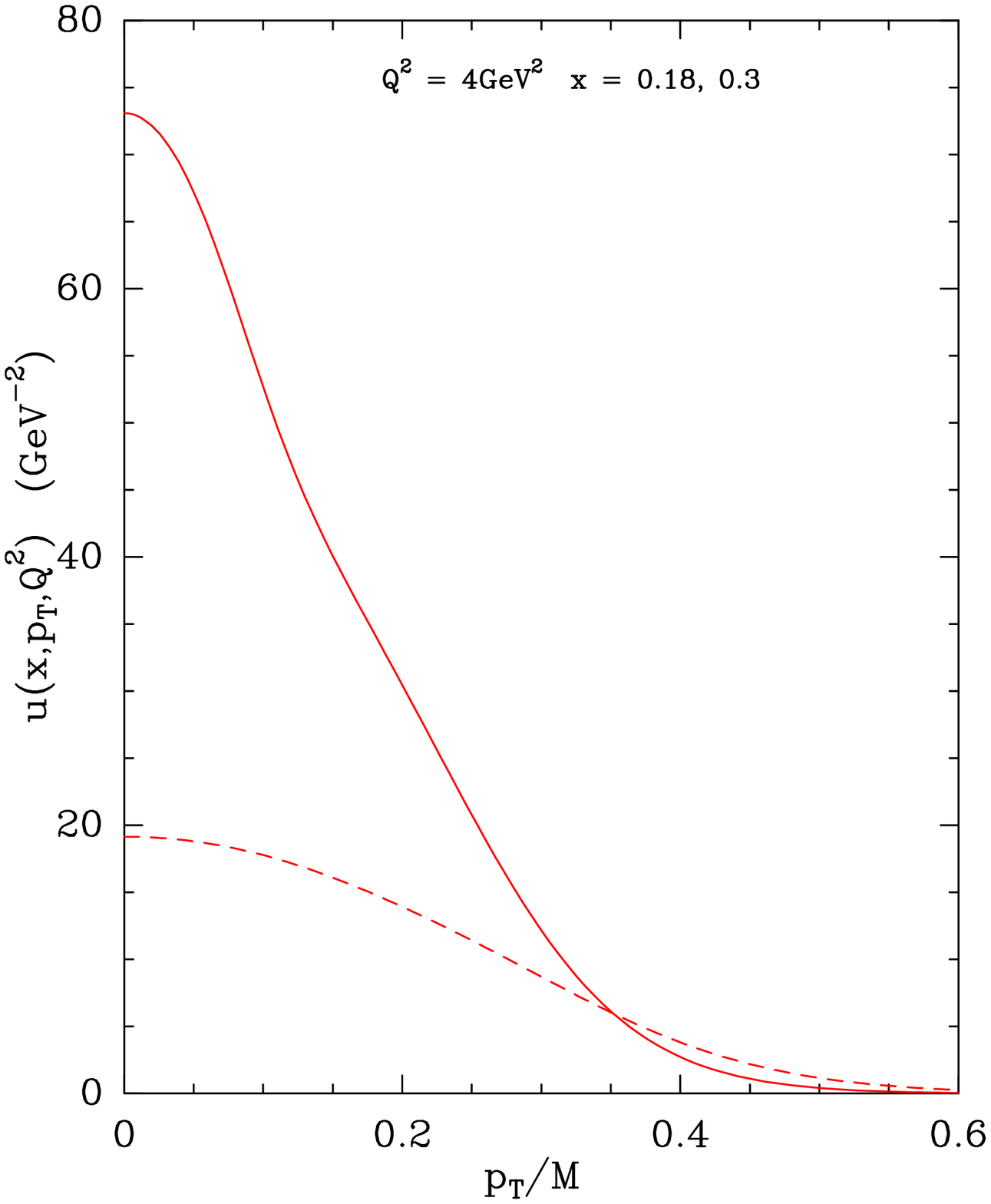,width=6.5cm}
  \end{minipage}
    \begin{minipage}{6.5cm}
  \epsfig{figure=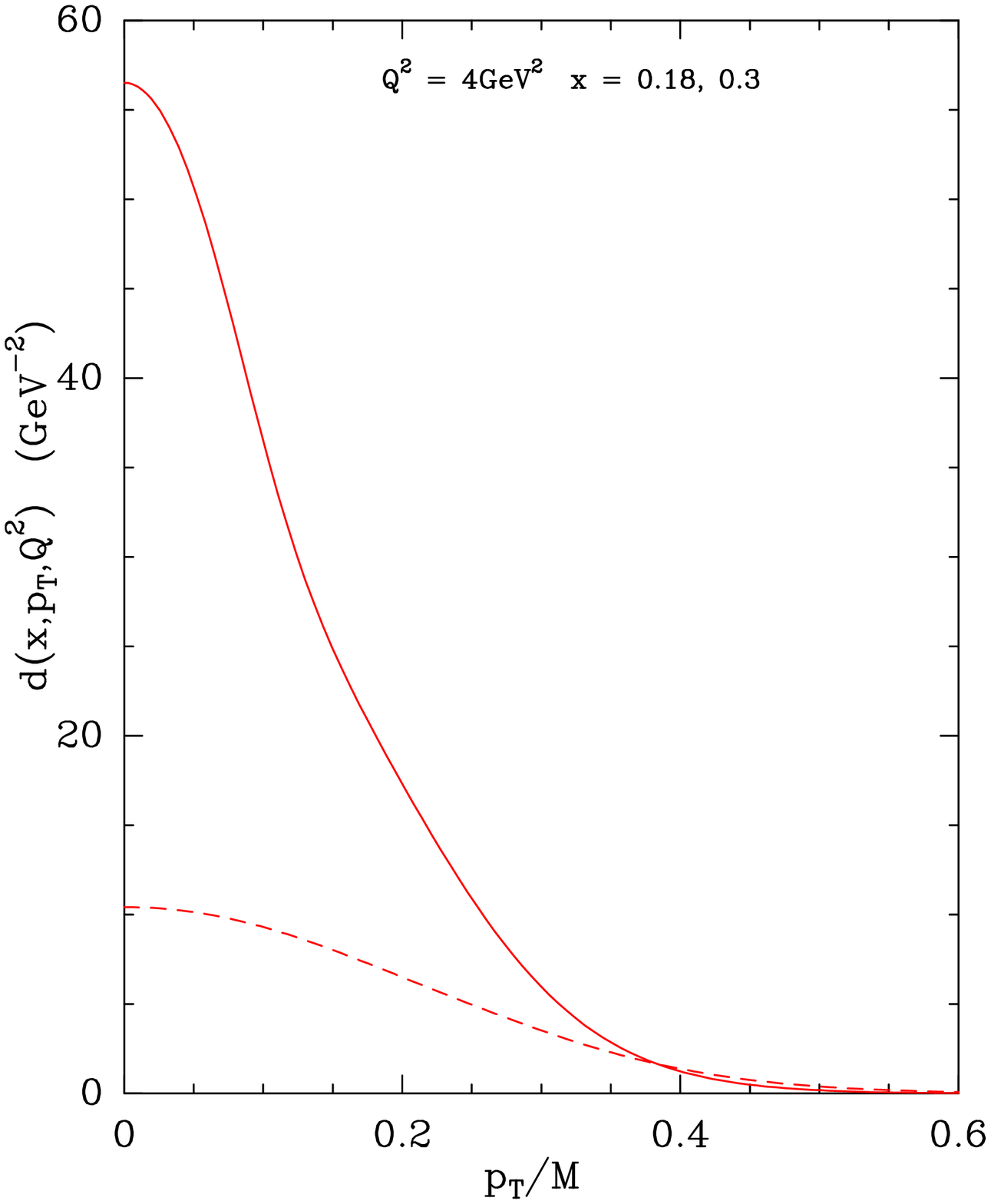,width=6.5cm}
    \end{minipage}
\end{center}
  \vspace*{-5mm}
\caption{The TMD distributions for $u$ and $d$ quarks, versus $p_T/M$, at $Q^2=4\mbox{GeV}^2$, for two $x$ values: solid lines $x=0.18$, dashed lines $x=0.3$.}
\label{comparison}
\vspace*{-1.5ex}
\end{figure}
\begin{figure}[ht]
\begin{center}
  \begin{minipage}{6.5cm}
  \epsfig{figure=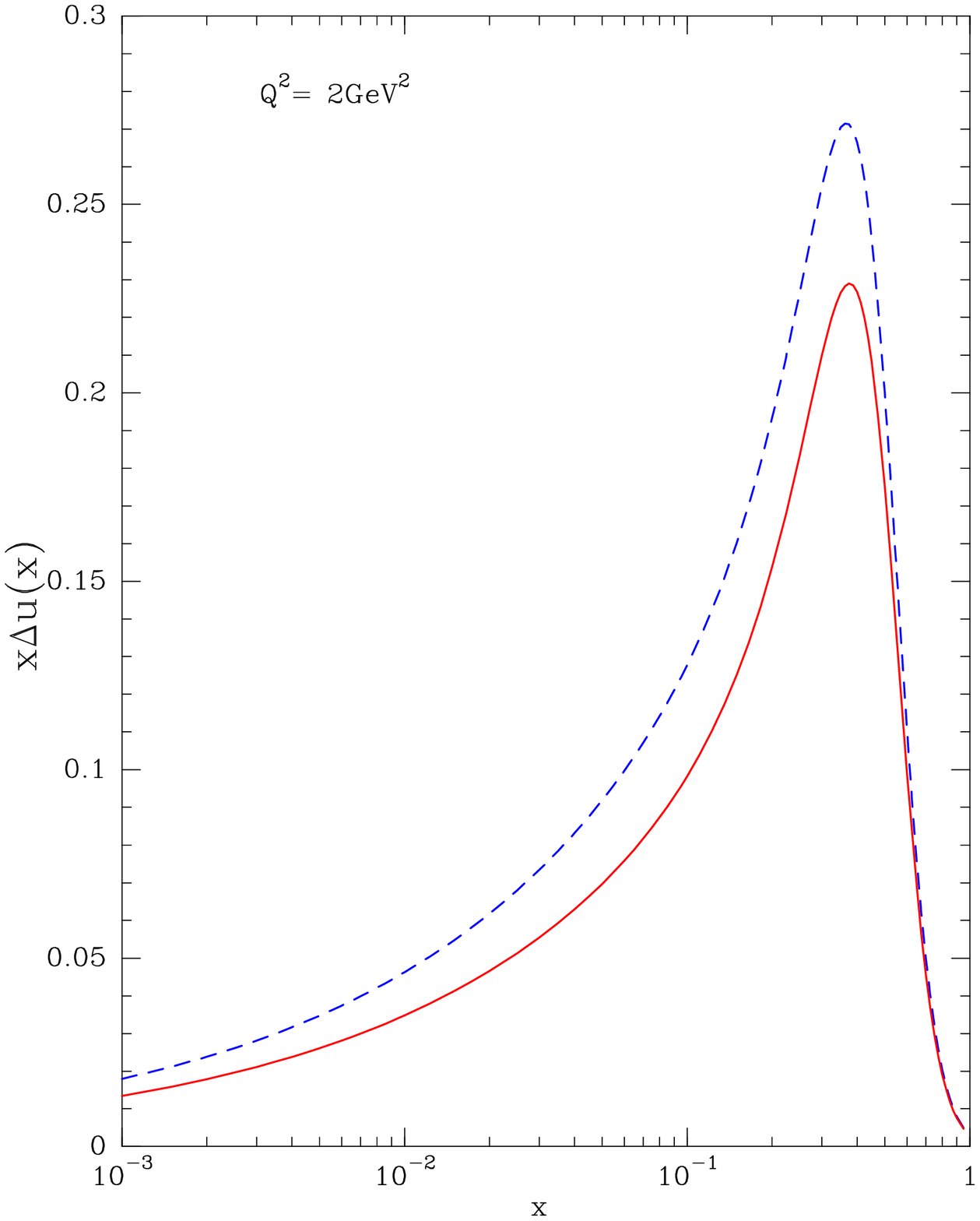,width=5.8cm}
  \end{minipage}
    \begin{minipage}{6.5cm}
  \epsfig{figure=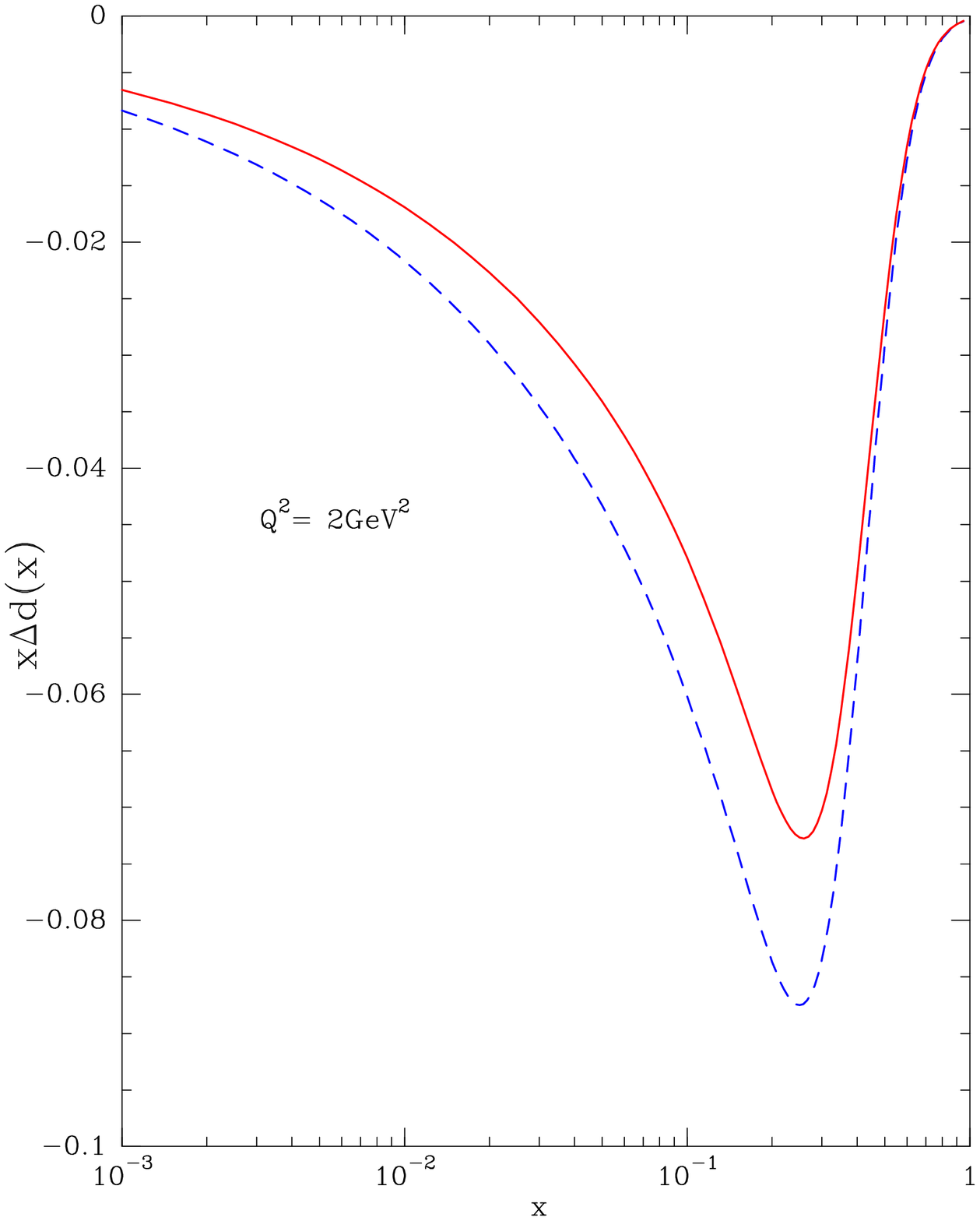,width=5.8cm}
    \end{minipage}
\end{center}
  \vspace*{-5mm}
\caption{The $u$ and $d$ quark helicity distributions versus $x$: $x\Delta q(x)$ ({\it dashed line}) and $x\Delta q^{MW}(x)$ ({\it solid line})}
\label{dsigcross1}
\vspace*{-1.5ex}
\end{figure}
The TMD helicity distributions $x\Delta q(x,p_T^2,P_z)$ will be defined as above, by substracting instead of adding the two helicity components, so there is no diffractive contribution.\\
It is clear that after integration over $p_T$, $\Delta q(x)$ does not depend on $P_z$.
However this is not the case for the helicity distributions modified by the Melosh-Wigner transformation, since after
taking into account the cancellation obtained in Section 3, we have
\beq
&&x\Delta q^{MW}(x,P_z) = \no\\
&&\frac{1}
{\mbox{exp}[(x - X_{0q}^{+})/{\bar x}] +1}F_q^+(x,P_z,\mu^2)
\mu^2 x\ln{(1+e^{Y_{0q}^+})} \no\\
&&- \frac{1}
{\mbox{exp}[(x - X_{0q}^{-})/{\bar x}] +1}F^-(x,P_z,\mu^2)
\mu^2 x\ln{(1+e^{Y_{0q}^-})} \label{eq116}~.
\eeq
By using Eq.(\ref{eq110}) one finds
\beq
x\Delta q^{MW}(x,P_z) &=& 
\frac{Ax^b}
{[\mbox{exp}[(x - X_{0q}^{+})/{\bar x}] +1][ 1 + R_q^+]} \no\\
&&-\frac{Ax^b}
{[\mbox{exp}[(x - X_{0q}^{-})/{\bar x}] +1][ 1 + R_q^-]} \label{117}~.
\eeq
As expected, in the limit $P_z \to \infty$, the Melosh-Wigner transformation
becomes an identity, so $x\Delta q^{MW}(x,P_z) \to x\Delta q(x)$. For illustration we display in Fig. 5, $x\Delta q(x)$
and $x\Delta q^{MW}(x)$ for $Q^2= 2\mbox{GeV}^2$, which shows the effect of the Melosh-Wigner rotation, mainly in the low $x$ region.

It is interesting to note that $|\Delta q^{MW}(x)| < |\Delta q(x)|$, as expected from some earlier work \cite{MSS}.
\section{Concluding remarks}
We have presented the extension of the statistical parton distributions to include
their transverse momentum dependence, by considering that the proton target has a finite longitudinal momentum $P_z$. This situation generates
some correction factors, which are only relevant in the small $Q^2$ region and for rather low $x$ values. The TMD distributions were generated by 
means of a transverse energy sum rule which implies that a simplifying factorization assumption is not allowed, as explained above. This sum rule has been used by other authors \cite{ZSM} who have, as in the present work, non factorized TMD distributions, as well as in the relativistic covariant approach \cite{ESTZ}. The TMD diffractive part of the quark (antiquark) distributions $q^{D}(x,p_T^2)$ and the TMD gluon distribution $G(x,p_T^2)$ had to be treated in a different way, as in Ref.\cite{BBS5}, in order to avoid a singularity in the energy sum rule. As a result their contributions to the sum rule are very small and the nondiffractive contributions dominate. Our approach involves the parameter $\mu^2$, which plays the role of the temperature for the transverse degrees of freedom, whose value is determined by the sum rule, as given in Ref.\cite{BBS5} which is probably an upper bound. If the TMD gluon distribution contributes significantly to the energy sum rule, as it does
 to the momentum sum rule, we might obtain a much smaller value, but this remains to be proven and will be the subject of a future work.

We have also shown the importance of the remarkable Melosh-Wigner transformation, whose effects are significant only for a finite $P_z$.

At the difference of other approaches, where statistical concepts are used
in the target rest frame and then a boost is applied to a large $P_z$ frame,
we made the choice to consider directly the large $P_z$ frame and the
$x$ variable to define $X_q^h$ and the shapes of the different distributions..

There is a phenomenological evidence that the partons dominating the large $x$
regions are not the same, as those which dominate the low $x$ region, as one could
mainly find by boosting an isotropical rest frame distributions. So we think that there is a deep theoretical reason to 
settle the statistical concepts directly in the $x$ variable related to the 
foundation of statistical mechanics. One may also cast some doubts on the use of the statistical approach, since the 
total number of valence partons is small, 2-$u$ quarks and 1-$d$ quark in the 
proton, but the fact that one writes a probability density makes reasonable to
apply a statistical approach to probabilities.\\
Finally, we would like to refer to an interesting recent paper suggesting a duality principle between our approach and a thermal description
of the PDF \cite{CLST}. This duality allows them to introduce an {\it effective temperature} T$\sim$120-150MeV, which is approximately the same for 
longitudinal and transverse momentum. The comparison with our results may be done in the Boltzmann limit, where we neglect the effect of
quantum statistics, which is crucial to get the phenomenological successful shape-first moment correlation for the valence partons and the isospin
and spin asymmetries of the sea. In that limit we get for the longitudinal temperature $T_l = M\bar{x}/2$, which gives 47MeV. To get the transverse temperature $T_t$, always in the Boltzmann limit, one should follow the method described in Ref.\cite{BP} leading to $T_t=\mu\sqrt{\bar{x}}/2$, which gives 70MeV, not too far from $T_l$. As noted in Ref.\cite{CLST}, there is no contradiction in getting different values in our approach and in the thermal description of the PDF.

\appendix
\section{Appendix}
\setcounter{equation}{0}
\numberwithin{equation}{section}

Let us recall that the nondiffractive contribution of the quark distribution $q^{h}(x)$ introduced earlier, must be obtained by an integration
over $p_T^2$, as follows
\be
 xq^h(x) =\int_0^\infty xq^h(x,p_T^2,P_z)dp_T^2\,,
\label{eq22}
\ee
where $xq^h(x,p_T^2,P_z)$ is given by Eq.(\ref{eqF}). So if we introduce the variable  $\xi=p_T^2/x\mu^{2}$,
we have to consider the following integral
\be
\int_{0}^{\infty} \frac{x\mu^2d\xi}{\exp{[\frac{2 \xi}
{1 + \sqrt{1 + \mu^2\xi/x P_z^2}} - Y_{0q}^h]} +1}\,.
\label{eq23}
\ee
A simple inspection of Eq.(\ref{eq23}) shows that the $p_T$ dependence is a 
complicated expression which involves a square root,
so the integral is certainly not tractable analytically. One way to by-pass
this difficulty is to transform the integrant in the form of a usual
Fermi-Dirac function and consequently to compute the associated differential
element.
In the above Eq.(\ref{eq23}) let us perform the change of variable 
\be
\eta = \frac{2\xi}{\sqrt{1 +\frac{\mu^2\xi}{x P_z^2}} +1} \,,
\label{eq27a}
\ee  
which can be written as
\be
\frac{2\xi}{\eta} = \sqrt{1 +\frac{\mu^2 \xi}{xP_z^2}} + 1\,.
\label{eq28}
\ee
Now from Eq.(\ref{eq28}) one has
\beq
1 +\frac{\mu^2\xi}{xP_z^2} &=& \frac{4\xi^2}{\eta^2} -\frac{4\xi}{\eta} +1
\label{eq31}\\
\frac{\mu^2}{xP_z^2} &=& \frac{4}{\eta}(\frac{\xi}{\eta} -1) \,,
\label{eq32}
\eeq
and finally the relation between $\eta$ and $\xi$
\be
\xi = \eta + \frac{\mu^2 \eta^2}{4 x P_z^2}\,.
\label{xidef}
\ee
By differentiation one gets
\be
d\xi = (1 +\frac{\mu^2 \eta}{2 x P_z^2}) d\eta \,,
\label{eq35a}
\ee
so with the above transformation, the integral (\ref{eq23}) can be rewritten
in a simplified form, close to a Fermi-Dirac distribution,
\be
\int_{0}^{\infty} \frac{x\mu^2 (1 +\frac{\mu^2 \eta}{2 x P_z^2})  d\eta}
{\mbox{exp}[\eta - Y_{0q}^h] +1}
\,.
\label{eq222}
\ee
To summarize we have shown that
\beq
&&\int_0^\infty\frac{dp^2_{T}}
{\mbox{exp}[2p^2_{T}/
[\mu^2(x + \sqrt{x^2 + p_T^2/P_z^2} )] - Y_{0q}^h]+1} \no \\
&&=\int_0^{\infty} \frac{x\mu^2  (1 +\frac{\mu^2 \eta}{2 x P_z^2})   d\eta}
{\mbox{exp}[ \eta - Y_{0q}^h] +1}=\mu^2 x[\ln{(1+e^{Y_{0q}^h})}
-\frac{\mu^2}{2xP_z^2} \mbox{Li}_2(-e^{Y_{0q}^h})]\no \\
&&=\mu^2 x\ln{(1\!+\!e^{Y_{0q}^h})}[ 1 \!+\! R_q^h]\,,
\label{eq112}
\eeq
where $\mbox{Li}_2$ denotes the polylogarithm function of order 2 and $R_q^h$ is a correction factor $R_q^h = - \frac{\mu^2}{2xP_z^2} \frac{\mbox{Li}_2(-e^{Y_{0q}^h})}{\ln{(1+e^{Y_{0q}^h})}}$, which involves the $P_z$ dependence.\\
We see that in the limit $P_z \rightarrow \infty$, we recover the
original formula given in Ref.\cite{BBS5}, for the integral of the TMD, since the correction factor disappears. 
In DIS, one can neglect $xM^2$ with respect to $Q^2$, so $P_z^2 = Q^2/4x(1-x)$ and 
\beq
R_q^h = - \frac{2\mu^2(1-x)}{Q^2} \frac{\mbox{Li}_2(-e^{Y_{0q}^h})}{\ln{(1+e^{Y_{0q}^h})}}~,
\label{eqR}
\eeq
and we note that $R_q^h >0$, because $\mbox{Li}_2(-e^{Y_{0q}^h})<0$.


\begin{thebibliography}{99}

\bibitem{BBS1} C. Bourrely, F. Buccella and J. Soffer, Eur. Phys. J. C {\bf 23}, 487 (2002);\\
 C. Bourrely, F. Buccella and J. Soffer, Mod. Phys. Lett. A {\bf 18}, 771 (2003)

\bibitem{Gott} K. Gottfried, Phys. Rev. Lett. {\bf 18}, 1154 (1967)

\bibitem{NMC} New Muon Collaboration, M. Arneodo {\it et al.}, Phys.
Rev. D {\bf 50}, R1 (1994) and references therein; 
P. Amaudruz {\it et al.}, Phys. Rev. Lett. {\bf 66}, 2712 (1991);
Nucl. Phys. B {\bf 371}, 3 (1992)


\bibitem{Pauli} A. Niegawa and K. Sasaki, Prog. Theo. Phys. {\bf 54}, 192 
(1975);\\ R.D. Field and R.P. Feynman, Phys. Rev. D {\bf 15}, 2590 (1977)

\bibitem{DY} R.S. Towell {\it et al.}, Phys. Rev. D {\bf 64}, 052002 (2001) and references therein

\bibitem{SVS} T. Sloan, R. Voss and G. Smadja, Phys. Rept. {\bf 162}, 45 (1988)

\bibitem{BBS3} C. Bourrely, F. Buccella and J. Soffer, Eur. Phys. J. C {\bf
41}, 327 (2005) 

\bibitem{BJ} J.D. Bjorken, Phys. Rev. D {\bf 1},1376 (1970)

\bibitem{hermes} HERMES Collaboration, X. Zheng {\it et al.}, Phys. Rev. C {\bf 70}, 065207 (2004)

\bibitem{JLab} Jefferson Lab Hall A Collaboration, A. Airapetian {\it et al.}, Phys. Rev. Lett. {\bf 92}, 012005 (2004)

\bibitem{BBS4} C. Bourrely, F. Buccella and J. Soffer, Mod. Phys. Lett. A
{\bf 21}, 143 (2006)

\bibitem{BBS5} C. Bourrely, F. Buccella and J. Soffer, Phys. Rev.
D {\bf 83}, 074003 (2011)

\bibitem{mel-wig} H.J. Melosh, Phys. Rev. D {\bf 9}, 1095 (1974); E. Wigner, Ann.
Math. {\bf 40}, 149 (1939)

\bibitem{bucc} F. Buccella, E. Celeghini, H. Kleinert, C.A. Savoy and E.
Sorace, Nuovo Cimento A {\bf 69}, 133 (1970); F. Buccella, C.A. Savoy and P. Sorba,
 Lettere al Nuovo Cimento {\bf 10}, 455 (1974)

\bibitem{MZ} B.-Q. Ma and Q.-R. Zhang, Z. Phys. C {\bf 58}, 479 (1993) and references therein

\bibitem{MSS} B.-Q. Ma, I. Schmidt and J. Soffer, Phys. Lett. B {\bf 441}, 461 (1998)

\bibitem{ESTZ} A.V. Efremov, P. Schweitzer, O.V. Teryaev and P. Zavada, Phys. Rev.
D {\bf 83}, 054025 (2011)

 \bibitem{ZSM} Y. Zhang, L. Shao and B.-Q. Ma, Phys. Lett. B {\bf
671}, 30 (2009)

 \bibitem{CLST} J. Cleymans, G.I. Lykasov, A.S. Sorin and O.V. Teryaev, Phys. Atom. Nucl. {\bf
75}, 725 (2012)

\bibitem{BP} F. Buccella and L. Popova, Mod. Phys. Lett. A
{\bf 17}, 2627 (2002)

\end{thebibliography}
\end{document}